\catcode`\@=11					



\font\fiverm=cmr5				
\font\fivemi=cmmi5				
\font\fivesy=cmsy5				
\font\fivebf=cmbx5				

\skewchar\fivemi='177
\skewchar\fivesy='60


\font\sixrm=cmr6				
\font\sixi=cmmi6				
\font\sixsy=cmsy6				
\font\sixbf=cmbx6				

\skewchar\sixi='177
\skewchar\sixsy='60


\font\sevenrm=cmr7				
\font\seveni=cmmi7				
\font\sevensy=cmsy7				
\font\sevenit=cmti7				
\font\sevenbf=cmbx7				

\skewchar\seveni='177
\skewchar\sevensy='60


\font\eightrm=cmr8				
\font\eighti=cmmi8				
\font\eightsy=cmsy8				
\font\eightit=cmti8				
\font\eightbf=cmbx8				

\skewchar\eighti='177
\skewchar\eightsy='60


\font\ninei=cmmi9
\font\ninesy=cmsy9

\skewchar\ninei='177
\skewchar\ninesy='60


\font\tenrm=cmr10				
\font\teni=cmmi10				
\font\tensy=cmsy10				
\font\tenex=cmex10				
\font\tenit=cmti10				
\font\tensl=cmsl10				
\font\tenbf=cmbx10				
\font\tentt=cmtt10				
\font\tenss=cmss10				
\font\tensc=cmcsc10				
\font\tenbi=cmmib10				

\skewchar\teni='177
\skewchar\tenbi='177
\skewchar\tensy='60

\def\tenpoint{\ifmmode\err@badsizechange\else
	\textfont0=\tenrm \scriptfont0=\sevenrm \scriptscriptfont0=\fiverm
	\textfont1=\teni  \scriptfont1=\seveni  \scriptscriptfont1=\fivemi
	\textfont2=\tensy \scriptfont2=\sevensy \scriptscriptfont2=\fivesy
	\textfont3=\tenex \scriptfont3=\tenex   \scriptscriptfont3=\tenex
	\textfont4=\tenit \scriptfont4=\sevenit \scriptscriptfont4=\sevenit
	\textfont5=\tensl
	\textfont6=\tenbf \scriptfont6=\sevenbf \scriptscriptfont6=\fivebf
	\textfont7=\tentt
	\textfont8=\tenbi \scriptfont8=\seveni  \scriptscriptfont8=\fivemi
	\def\rm{\tenrm\fam=0 }%
	\def\it{\tenit\fam=4 }%
	\def\sl{\tensl\fam=5 }%
	\def\bf{\tenbf\fam=6 }%
	\def\tt{\tentt\fam=7 }%
	\def\ss{\tenss}%
	\def\sc{\tensc}%
	\def\bmit{\fam=8 }%
	\rm\setparameters\setbaselines\fi}


\font\twelverm=cmr12				
\font\twelvei=cmmi12				
\font\twelvesy=cmsy10	scaled\magstep1		
\font\twelveex=cmex10	scaled\magstep1		
\font\twelveit=cmti12				
\font\twelvesl=cmsl12				
\font\twelvebf=cmbx12				
\font\twelvett=cmtt12				
\font\twelvess=cmss12				
\font\twelvesc=cmcsc10	scaled\magstep1		
\font\twelvebi=cmmib10	scaled\magstep1		

\skewchar\twelvei='177
\skewchar\twelvebi='177
\skewchar\twelvesy='60

\def\twelvepoint{\ifmmode\err@badsizechange\else
	\textfont0=\twelverm \scriptfont0=\eightrm \scriptscriptfont0=\sixrm
	\textfont1=\twelvei  \scriptfont1=\eighti  \scriptscriptfont1=\sixi
	\textfont2=\twelvesy \scriptfont2=\eightsy \scriptscriptfont2=\sixsy
	\textfont3=\twelveex \scriptfont3=\tenex   \scriptscriptfont3=\tenex
	\textfont4=\twelveit \scriptfont4=\eightit \scriptscriptfont4=\sevenit
	\textfont5=\twelvesl
	\textfont6=\twelvebf \scriptfont6=\eightbf \scriptscriptfont6=\sixbf
	\textfont7=\twelvett
	\textfont8=\twelvebi \scriptfont8=\eighti  \scriptscriptfont8=\sixi
	\def\rm{\twelverm\fam=0 }%
	\def\it{\twelveit\fam=4 }%
	\def\sl{\twelvesl\fam=5 }%
	\def\bf{\twelvebf\fam=6 }%
	\def\tt{\twelvett\fam=7 }%
	\def\ss{\twelvess}%
	\def\sc{\twelvesc}%
	\def\bmit{\fam=8 }%
	\rm\setparameters\setbaselines\fi}


\font\fourteenrm=cmr12	scaled\magstep1		
\font\fourteeni=cmmi12	scaled\magstep1		
\font\fourteensy=cmsy10	scaled\magstep2		
\font\fourteenex=cmex10	scaled\magstep2		
\font\fourteenit=cmti12	scaled\magstep1		
\font\fourteensl=cmsl12	scaled\magstep1		
\font\fourteenbf=cmbx12	scaled\magstep1		
\font\fourteentt=cmtt12	scaled\magstep1		
\font\fourteenss=cmss12	scaled\magstep1		
\font\fourteensc=cmcsc10 scaled\magstep2	
\font\fourteenbi=cmmib10 scaled\magstep2	

\skewchar\fourteeni='177
\skewchar\fourteenbi='177
\skewchar\fourteensy='60

\def\fourteenpoint{\ifmmode\err@badsizechange\else
	\textfont0=\fourteenrm \scriptfont0=\tenrm \scriptscriptfont0=\sevenrm
	\textfont1=\fourteeni  \scriptfont1=\teni  \scriptscriptfont1=\seveni
	\textfont2=\fourteensy \scriptfont2=\tensy \scriptscriptfont2=\sevensy
	\textfont3=\fourteenex \scriptfont3=\tenex \scriptscriptfont3=\tenex
	\textfont4=\fourteenit \scriptfont4=\tenit \scriptscriptfont4=\sevenit
	\textfont5=\fourteensl
	\textfont6=\fourteenbf \scriptfont6=\tenbf \scriptscriptfont6=\sevenbf
	\textfont7=\fourteentt
	\textfont8=\fourteenbi \scriptfont8=\tenbi \scriptscriptfont8=\seveni
	\def\rm{\fourteenrm\fam=0 }%
	\def\it{\fourteenit\fam=4 }%
	\def\sl{\fourteensl\fam=5 }%
	\def\bf{\fourteenbf\fam=6 }%
	\def\tt{\fourteentt\fam=7}%
	\def\ss{\fourteenss}%
	\def\sc{\fourteensc}%
	\def\bmit{\fam=8 }%
	\rm\setparameters\setbaselines\fi}


\font\seventeenrm=cmr10 scaled\magstep3		


\newdimen\rp@
\newcount\@basestretchnum
\newskip\@baseskip
\newskip\headskip
\newskip\footskip


\def\setparameters{\rp@=.1em
	\headskip=24\rp@
	\footskip=\headskip
	\delimitershortfall=5\rp@
	\nulldelimiterspace=1.2\rp@
	\scriptspace=0.5\rp@
	\abovedisplayskip=10\rp@ plus3\rp@ minus5\rp@
	\belowdisplayskip=10\rp@ plus3\rp@ minus5\rp@
	\abovedisplayshortskip=5\rp@ plus2\rp@ minus4\rp@
	\belowdisplayshortskip=10\rp@ plus3\rp@ minus5\rp@
	\normallineskip=\rp@
	\lineskip=\normallineskip
	\normallineskiplimit=0pt
	\lineskiplimit=\normallineskiplimit
	\jot=3\rp@
	\setbox0=\hbox{\the\textfont3 B}\p@renwd=\wd0
	\skip\footins=12\rp@ plus3\rp@ minus3\rp@
	\skip\topins=0pt plus0pt minus0pt}


\def\setbaselines{\maxdepth=4\rp@\baselinestretch=\@basestretchnum}


\def\baselinestretch{\afterassignment\@basestretch\@basestretchnum}
\def\@basestretch{%
	\@baseskip=12\rp@ \divide\@baseskip by1000
	\normalbaselineskip=\@basestretchnum\@baseskip
	\baselineskip=\normalbaselineskip
	\bigskipamount=\the\baselineskip
		plus.25\baselineskip minus.25\baselineskip
	\medskipamount=.5\baselineskip
		plus.125\baselineskip minus.125\baselineskip
	\smallskipamount=.25\baselineskip
		plus.0625\baselineskip minus.0625\baselineskip
	\setbox\strutbox=\hbox{\vrule height.708\baselineskip
		depth.292\baselineskip width0pt }}



\def\makeheadline{\vbox to0pt{\baselinestretch=1000
	\vskip-\headskip \vskip1.5pt
	\line{\vbox to\ht\strutbox{}\the\headline}\vss}\nointerlineskip}

\def\makefootline{\baselineskip=\footskip\line{\the\footline}}

\def\big#1{{\hbox{$\left#1\vbox to8.5\rp@ {}\right.\n@space$}}}
\def\Big#1{{\hbox{$\left#1\vbox to11.5\rp@ {}\right.\n@space$}}}
\def\bigg#1{{\hbox{$\left#1\vbox to14.5\rp@ {}\right.\n@space$}}}
\def\Bigg#1{{\hbox{$\left#1\vbox to17.5\rp@ {}\right.\n@space$}}}


\mathchardef\alpha="710B
\mathchardef\beta="710C
\mathchardef\gamma="710D
\mathchardef\delta="710E
\mathchardef\epsilon="710F
\mathchardef\zeta="7110
\mathchardef\eta="7111
\mathchardef\theta="7112
\mathchardef\iota="7113
\mathchardef\kappa="7114
\mathchardef\lambda="7115
\mathchardef\mu="7116
\mathchardef\nu="7117
\mathchardef\xi="7118
\mathchardef\pi="7119
\mathchardef\rho="711A
\mathchardef\sigma="711B
\mathchardef\tau="711C
\mathchardef\upsilon="711D
\mathchardef\phi="711E
\mathchardef\chi="711F
\mathchardef\psi="7120
\mathchardef\omega="7121
\mathchardef\varepsilon="7122
\mathchardef\vartheta="7123
\mathchardef\varpi="7124
\mathchardef\varrho="7125
\mathchardef\varsigma="7126
\mathchardef\varphi="7127
\mathchardef\imath="717B
\mathchardef\jmath="717C
\mathchardef\ell="7160
\mathchardef\wp="717D
\mathchardef\partial="7140
\mathchardef\flat="715B
\mathchardef\natural="715C
\mathchardef\sharp="715D


\def\err@badsizechange{%
	\immediate\write16{--> Size change not allowed in math mode, ignored}}

\baselinestretch=1000
\tenpoint

\catcode`\@=12					
\catcode`\@=11
\expandafter\ifx\csname @iasmacros\endcsname\relax
	\global\let\@iasmacros=\par
\else	\immediate\write16{}
	\immediate\write16{Warning:}
	\immediate\write16{You have tried to input iasmacros more than once.}
	\immediate\write16{}
	\endinput
\fi
\catcode`\@=12


\def\rmb{\seventeenrm}

\def\singlespace{\baselineskip=\normalbaselineskip}
\def\halfspace{\baselineskip=1.5\normalbaselineskip}
\def\doublespace{\baselineskip=2\normalbaselineskip}


\def\AB{\bigskip\parindent=40pt
        \centerline{\bf ABSTRACT}\medskip\halfspace\narrower}
\def\AE{\bigskip\nonarrower\doublespace}
\def\nonarrower{\advance\leftskip by-\parindent
	\advance\rightskip by-\parindent}


\def\boxit#1{\vbox{\hrule\hbox{\vrule\kern3pt
	\vbox{\kern3pt#1\kern3pt}\kern3pt\vrule}\hrule}}

\def\hence{\leavevmode\hbox{\bf .\raise5.5pt\hbox{.}.} }

\def\dalemb#1#2{{\vbox{\hrule height.#2pt
	\hbox{\vrule width.#2pt height#1pt \kern#1pt \vrule width.#2pt}
	\hrule height.#2pt}}}
\def\gtorder{\mathrel{\raise.3ex\hbox{$>$}\mkern-14mu
             \lower0.6ex\hbox{$\sim$}}}
\def\ltorder{\mathrel{\raise.3ex\hbox{$<$}\mkern-14mu
             \lower0.6ex\hbox{$\sim$}}}

\newdimen\fullhsize
\newbox\leftcolumn
\def\twoup{\hoffset=-.5in \voffset=-.25in
  \hsize=4.75in \fullhsize=10in \vsize=6.9in
  \def\fullline{\hbox to\fullhsize}
  \let\lr=L
  \output={\if L\lr
        \global\setbox\leftcolumn=\columnbox\global\let\lr=R \advancepageno
      \else \doubleformat \global\let\lr=L\fi
    \ifnum\outputpenalty>-20000 \else\dosupereject\fi}
  \def\doubleformat{\shipout\vbox{
    \fullline{\box\leftcolumn\hfil\columnbox}\advancepageno}}
  \def\columnbox{\leftline{\vbox{\makeheadline\pagebody\makefootline}}}
  \tolerance=1000 }
\overfullrule=0pt
\twelvepoint
\doublespace
{\nopagenumbers{
\rightline{~~~March, 2002}
\bigskip\bigskip
\centerline{\rmb Why Decoherence has not Solved the Measurement Problem:}
\centerline{\rmb A Response to P.W. Anderson}
\medskip
\centerline{\it  Stephen L. Adler
}
\centerline{\bf Institute for Advanced Study}
\centerline{\bf Princeton, NJ 08540}
\medskip
\bigskip\bigskip
\leftline{\it Send correspondence to:}
\medskip
{\singlespace\leftline{Stephen L. Adler}
\leftline{Institute for Advanced Study}
\leftline{Einstein Drive, Princeton, NJ 08540}
\leftline{Phone 609-734-8051; FAX 609-924-8399; email adler@ias.edu}}
\bigskip\bigskip
}}
\vfill\eject
\pageno=2
\AB
We discuss why, contrary to claims recently made by P.W. Anderson,  
decoherence has not solved the quantum measurement problem.  
\bigskip
\bigskip
{\it Keywords:}  Decoherence; Measurement Problem; Collapse; 
Reduction
\AE
\vfill \eject
It has lately become fashionable to claim that decoherence has solved 
the quantum measurement problem by eliminating the necessity for
Von Neumann's wave function collapse postulate.  For example, in 
a recent review in 
{\it  Studies in History and Philosophy of
Modern Physics},  Anderson (2001) 
states ``The last chapter...
deals with the quantum measurement problem....My main test, allowing me to 
bypass the extensive discussion, was a quick, unsuccessful search in the 
index for the word `decoherence' which describes the process that used to 
be called `collapse of the wave function'.  The concept is now 
experimentally verified by beautiful atomic beam techniques quantifying the 
whole process.''  And again, in his response to the author's response  
(Anderson, 2001),  
``Our difference about `decoherence' is real.  I find this, and the cluster
of ideas around it much preferable to more traditional ways of treating 
the quantum paradoxes because the `classical' apparatus is treated as 
a quantum system as well....; and as I remarked, recent experiments have 
verified this approach.''  In a somewhat similar vein, Tegmark and Wheeler 
(2001) 
state in a recent {\it Scientific American} 
article discussing the ``many-worlds'' interpretation of quantum mechanics 
and decoherence, ``...it is time to update 
the quantum textbooks: although these infallibly list explicit non-unitary 
collapse as a fundamental postulate in one of the early chapters, ...many 
physicists ... no longer take this seriously.  The notion of collapse will 
undoubtedly retain great utility as a calculational recipe, but an added 
caveat clarifying that it is probably not a fundamental process violating 
Schr\"odinger's equation could save astute students many hours of frustrated 
confusion.''

These striking statements to the contrary, I do not believe that 
either detailed theoretical calculations or recent  
experimental results show that decoherence has 
resolved the difficulties associated with   
quantum measurement theory.  This will not be a surprise to 
many workers in the field of 
decoherence; for example, in their seminal paper on 
decoherence as a source 
of spatial localization, Joos and Zeh (1985) state 
``Of course no unitary treatment 
of the time dependence can explain why only one of these dynamically 
independent components is experienced.''  And in a recent review on 
decoherence,  Joos (1999) 
states ``Does decoherence solve the measurement problem?  Clearly not.  
What decoherence tells us is that certain objects {\it appear} classical 
when observed.  But what is an observation?  At some stage we still have to 
apply the usual probability rules of quantum theory.''  Going back a few 
years, an informative 
and lively debate on these issues can be found in the Letters column of 
the April 1993 {\it Physics Today} (starting on page 13 of that issue and 
continuing over many pages), in response to an earlier article in that 
journal by Zurek (1991). An enlightening discussion of the measurement 
problem has been given by Bell (1990), and there also are extensive discussions of both 
the measurement problem and the role of decoherence in the philosophy of 
physics literature.  A careful analysis of the measurement problem has 
been given by Brown (1986), who reviews earlier work of Fine (1969) 
and others.   Rebuttals to the claim that decoherence solves the measurement 
problem have been given in the books of Albert (1992), Barrett (1999) and 
Bub (1997), with Bub's treatment closet in spirit to the formulation given 
below.  A detailed analysis of decoherence within the consistent histories 
approach has been given by  Kent and McElwaine (1997), and 
 discussions of  decoherence in the context of the many-worlds approach  
  can be found in Bacciagaluppi (2001)
 (who  gives  an extensive bibliography on decoherence 
as it relates to the measurement problem) and in Butterfield (2001).   
Despite the existence of these and other prior  
discussions, I think it worthwhile to revisit the substantive issues, 
particularly in the light of recent claims that decoherence resolves  
the measurement problem.  

Let me begin by setting up a simple quantum mechanical 
model for measurement, in which the 
effects of decoherence can be explicitly taken into account.  
Let us consider 
a two state system $X$ with initial ($t=0$) state vector 
$$|\psi_0\rangle_X=\alpha |\psi^{(A)}\rangle_X+
\beta |\psi^{(B)}\rangle_X~~~,\eqno(1)$$ 
with $|\alpha|^2+|\beta|^2=1$, which could be,   
for example, an atomic ion (or nucleus) with up and down spin states and 
a high enough energy gap for internal excitations so that these can be 
neglected.  This system is  
probed by an apparatus with initial state vector $|\phi_0\rangle_{APP}$, 
and the system and apparatus interact in turn with an environment with 
initial state vector $|\phi_0\rangle_{ENV}$.  Thus, at the start of 
the measurement process, the total wave function is the direct product 
$$|\Phi_0\rangle =|\psi_0\rangle_X 
|\phi_0\rangle_{APP} |\phi_0\rangle_{ENV} ~~~.\eqno(2)$$

Before going on to discuss the time development of this system, let me  
comment on assumptions that are implicit in Eq.~(2).  First of all, there 
is nothing subjective about the separation into system, apparatus, and 
environment.  At the low energies characterizing atomic beam experiments,  
atomic nuclei are conserved species of particles, with an independent 
conservation law for each atomic number and for each isotope.  Hence 
with a system consisting, say, of a silver ion or nucleus, an apparatus 
constructed from iron, copper, etc., and an environment consisting 
of the atmospheric gases together with  
electromagnetic radiation, there is an 
objective distinction between system, apparatus, and environment, even   
when indistinguishability of like particles is taken into account.  
Secondly, convergent calculations presuppose a finite dimensional Hilbert 
space (unless careful precautions are taken).  Once high energy or short 
distance physics is subsumed in renormalized parameters, 
this assumption is obeyed for the system and apparatus, which contain 
a finite number 
of atoms.  How many particles must be considered in the environment?  
Letting $T$ be the time allotted for the measurement (typically of order 
$10^{-3}$  seconds in an atomic beam experiment), only those 
environmental particles that are within a radius 
$R=cT$ of the apparatus, with $c$ the velocity of light,   
can causally influence the experimental outcome.  We shall use this criterion 
to define the number of particles in the environment; it is clearly  
tens of orders of magnitudes smaller than the number of particles in the 
visible universe.  While it may be questionable whether the Schr\"odinger 
equation can be applied to the entire universe, there should be no 
difficulty, if quantum mechanics is an exact theory, in 
applying the Schr\"odinger equation to the state vector for all 
environmental particles within the causal horizon $R$ for the apparatus.  

Let us now allow the initial state vector of Eq.~(2) to evolve in time 
according to the deterministic unitary evolution $U=\exp(-iHt)$.  In  
general, the two system states, the apparatus states, and the environment 
states become entangled in a generic way, so that all we can say is that 
the evolved wave function is $|\Phi(t)\rangle = U|\Phi_0\rangle$, which 
conveys no useful information.  However, for a cleverly devised apparatus, 
such as that used in a Stern-Gerlach or similar molecular beam 
experiment, we can guarantee 
that the state that evolves from the initial state $|\Phi_0\rangle$ 
has the form 
$$|\Phi(t)\rangle= \alpha |\psi^{(A)}\rangle_X  
|\phi^{(A)}(t)\rangle_{APP+ENV} 
+\beta |\psi^{(B)}\rangle_X |\phi^{(B)}(t)\rangle_{APP+ENV}~~~,\eqno(3)$$
where $|\phi^{(A)}(t)\rangle_{APP+ENV}$  
and $|\phi^{(B)}(t)\rangle_{APP+ENV}$ 
are entangled states of the apparatus and environment that by time $T$ are 
macroscopically distinguishable.  The presence or absence of  
$|\phi^{(A,B)}(T)\rangle_{APP+ENV}$ 
is registered by a recording device, which long after $T$ can be read by 
an observer who travels in from outside the causal horizon of the apparatus.  

Even in this generality, without invoking a specific model, we can state 
the effect of decoherence.  What decoherence does is to cause the 
rapid decay with time of the inner product 
$$_{APP+ENV}\langle \phi^{(A)}(t)|\phi^{(B)}(t)\rangle_{APP+ENV} ~~~,\eqno(4)$$  
which at time $t=0$ was unity.  As a consequence, interference effects  
between the system states $|\psi^{(A)}\rangle_X$ and $|\psi^{(B)}\rangle_X$, 
which are initially present, rapidly disappear as time evolves.   
A widely used model for decoherence, originated by  
Harris and Stodolsky (1981) in 
the context of optically active molecules, and applied to spatial 
localization of macroscopic objects by Joos and Zeh (1985), fits readily into 
this framework.  In this model one neglects the interactions of 
the environmental particles with each other and with the system $X$, 
and treats their scattering from the 
apparatus in an elastic, static, instantaneous approximation, so that there 
is no excitation of internal degrees of freedom of the apparatus,  
recoil of the apparatus is ignored, and the effect of scattering of a  
particle of the environment on the apparatus 
is described by the action of the 
single particle $S$ matrix on the particle initial state.  In this  
approximation, 
$$|\phi^{(J)}(t)\rangle_{APP+ENV} = |\phi^{(J)}(t)\rangle_{APP} 
\prod_i S_{i(J)}|\phi_0\rangle_{ENV,i} \prod_k |\phi_0\rangle_{ENV,k} 
~~,~~J=A,B,~~~\eqno(5a)$$
with $|\phi_0\rangle_{ENV,i}$ the initial state of the $i$th environmental 
particle that has scattered by time $t$, with $|\phi_0\rangle_{ENV,k}$ the 
initial state of the $k$th environmental particle that has {\it not} 
scattered by time $t$, and with $S_{i(J)}$ the 
$S$ matrix for the $i$th particle to scatter from the state $(J)$ of 
the apparatus.   (Thus the products over $i$ and $k$ together extend over 
all particles in the environment.)   In the approximation of Eq.~(5a), 
the inner product of Eq.~(4) becomes 
$${}_{APP+ENV}\langle \phi^{(A)}(t)|\phi^{(B)}(t)\rangle_{APP+ENV} 
\simeq {}_{APP}\langle \phi^{(A)}(t)|\phi^{(B)}(t) \rangle_{APP} 
\prod_i {}_{ENV,i}\langle  \phi_0| S^{\dagger}_{i(A)} S_{i(B)} 
|\phi_0\rangle_{ENV,i}
~~~.\eqno(5b)$$ 
Since each factor 
${}_{ENV,i}\langle  \phi_0| S^{\dagger}_{i(A)} S_{i(B)} 
|\phi_0\rangle_{ENV,i}$ 
is in general 
of magnitude less than 1, and since the number of factors in the product 
over $i$ grows approximately linearly with the elapsed time, the off 
diagonal matrix element of Eq.~(5b) approaches zero as $\exp(-\Lambda t)$, 
with $\Lambda$ defining the decoherence rate, as $t$ becomes large.  This 
formalism has been applied to estimate decoherence rates in a large variety 
of processes of physical interest, and as we have already noted, fits 
seamlessly into the framework of our more general discussion. 

Returning to the general formula of 
Eq.~(3), the quantum measurement problem consists in the observation that 
Eq.~(3) is {\it not} what is observed as the outcome of a measurement!
What is seen is  not the superposition of Eq.~(3),  
but rather {\it either} the unit normalized 
state 
$$|\psi^{(A)}\rangle_X  |\phi^{(A)}(t)\rangle_{APP+ENV}~~~,\eqno(6a)$$ 
{\it or} the unit normalized state
$$|\psi^{(B)}\rangle_X |\phi^{(B)}(t)\rangle_{APP+ENV}~~~.\eqno(6b)$$   
But because these 
states are orthogonal, they cannot both have evolved from a single 
initial state by a deterministic, unitary evolution, since 
it is a property of unitary transformations in Hilbert 
space that if $|(A)\rangle =
U|0\rangle$, and $|(B) \rangle =U |0\rangle$, then $\langle (A)|(B)\rangle =
\langle  0|U^{\dagger}U| 0 \rangle=1$.  Thus, when quantum mechanics is 
applied uniformly at all levels, to the apparatus and its environment as 
well as to the system, we are faced with a contradiction.  This 
contradiction is in no way ameliorated by decoherence, since the inner 
product of Eq.~(5b) plays no role in the final state vector of Eq.~(6a) or  
Eq.~(6b) 
that describes the outcome of the measurement.  
Note also that to see this contradiction 
we do not need an 
infinite sequence of repetitions of the experiment, as would be needed 
to discuss the probabilities of the outcomes $(A)$ and $(B)$, since only 
enough repetitions are needed to achieve an outcome $(A)$ and an outcome 
$(B)$ at least once.\footnote{$^1$}{The preceding argument is based on  
a simplified, ``bare bones'' statement of the measurement problem,  which 
suffices to demonstrate the ineffectiveness of decoherence.  
More precise treatments of the measurement problem can be found   
in the article of Brown (1986) and in the book of Bub (1997).  Additionally,  
the assumption of a deterministic 
unitary evolution can be weakened to that of a deterministic linear 
evolution, as discussed for example by Bassi and Ghirardi (2000).} 

What are the ways out of this dilemma?  One route, discussed in detail 
in the book of DeWitt and Graham (1973), is to insist that the superposition 
of Eq.~(3) {\it is} the final outcome of the measurement, not Eqs. (6a) or 
(6b), with the world state vector splitting into two branches, only one of 
which we observe.  A measure is then postulated on the space of world 
state vectors, with respect to which outcomes obeying the Born rule are 
typical, while outcomes not obeying the Born rule are of measure zero.   
A second route, represented 
by Bohmian quantum mechanics [Bohm (1952) and  D\"urr, Goldstein and 
Zanghi (1992)] and, alternatively, Ax-Kochen quantum mechanics 
[Ax and Kochen (1999)], 
supplements the usual quantum mechanical formalism with an associated 
configuration space, which plays a role in picking individual outcomes,   
in such a way that Born rule probabilities emerge for the observer who  
cannot access the additional information contained in the auxiliary  
configuration  space. These approaches are all {\it interpretations} of 
quantum mechanics, in the specific sense 
that by design they reproduce all physical predictions 
of quantum mechanics, and so are empirically indistinguishable 
from the orthodox theory, 
while changing the mathematical foundations so as to resolve the difficulties 
associated with measurement theory.    

If we insist on having only one world existing within the standard   
arena of states and operators in Hilbert space, we must instead 
discard one or more 
of the assumptions made in our analysis above, by injecting new physics.    
One alternative is to preserve 
the deterministic unitary evolution of quantum mechanics, and to 
drop the assumption that the environment and/or apparatus 
can always be prepared in a 
specified initial state. One might then attempt to show that the discrete 
choice of experimental outcome is tied to details of the  
initial state, giving a sense in which ``decoherence'', as understood   
more generally to mean environmental influence, 
could be said resolve the measurement 
problem.  A calculation showing how this might happen has never been 
given, and in fact, G. Gr\"ubl (2002) has recently pointed out that a 
modification of the generalized formulation of the measurement 
problem given by Bassi and Ghirardi (2000) shows, under very 
weak assumptions,  that initial state  
environmental effects cannot explain the occurrence of definite experimental  
outcomes.

Another alternative, which can also be formulated within the 
standard state vector and operator apparatus of Hilbert space, is to 
abandon the assumption of a deterministic unitary evolution, and to 
suppose instead that the evolution is stochastic unitary, in the sense 
that while the wave function for an individual system evolves unitarily, 
this evolution has a random, or stochastic component.    This approach  
was pioneered by Ghirardi, Rimini, and Weber (1986), 
Pearle (1976, 1979),  Gisin (1984) and Di'osi (1988), 
and has been studied by many others.\footnote{$^2$} 
{See   Adler, Brody, Brun and 
Hughston (2001) and Adler (2002) for recent mathematical     
and phenomenological analyses, respectively.  These papers give further 
references to the literature on the stochastic reduction approach.}  
It is implemented by 
replacing the standard Schr\"odinger evolution of the state vector by 
a stochastic differential evolution, in which $d|\psi\rangle$ receives a 
contribution proportional to $dt$ together with an 
additional small, nonlinear 
contribution proportional to $dW_t$, with  $dW_t$  a stochastic differential 
obeying $dW_t^2=dt$.  Heuristically, the idea here is that quantum mechanics 
may be modified by a low level universal noise, akin to Brownian motion, 
possibly arising from new physics at the Planck scale, which 
in certain situations causes reduction of the state vector.  
This approach, which has been developed in great detail, reproduces 
the observed fact of discrete outcomes governed by Born rule probabilities, 
and predicts the maintenance of coherence where that is observed (including 
in systems with large numbers of particles, such as recent superconductive 
tunneling and molecular diffraction experiments), while predicting 
state vector reduction when the apparatus parameters are those 
characterizing measurements.
The stochastic approach may ultimately be falsified 
by experiment, but it constitutes a viable phenomenological 
solution to the measurement problem.  Decoherence, in the 
absence of a detailed theory showing that   
it leads to stochastic outcomes with the correct properties, has yet 
to achieve this status.

\bigskip\bigskip
\vfill\eject
\bigskip
\centerline{\bf Acknowledgments}
This work was supported in part by the Department of Energy under
Grant \#DE--FG02--90ER40542.   I wish to thank Phillip Anderson, 
Todd Brun,  Jeremy Butterfield, Sheldon Goldstein, Gebhard Gr\"ubl, 
Ross Hyman, Leo Stodolsky, and the referees, for helpful comments in 
conversations or via email correspondence.     
\vfill\eject
\centerline{\bf References}
\bigskip
\noindent
Adler, S. L. (2002)  `Environmental Influence on the Measurement Process 
in Stochastic Reduction Models', {\it J. Phys. A: Math. Gen.} {\bf 35}, 
841-858.  
\bigskip 
\noindent
Adler, S. L., Brody, D. C., Brun, T. A. and Hughston, L. P. (2001) 
`Martingale models for quantum state reduction', 
{\it J. Phys. A: Math. Gen.}, 
{\bf 34}, 8795-8820. 
\bigskip
\noindent
Albert, D. Z. (1992) {\it Quantum Mechanics and Experience}, pp. 88-92 and  
161-164. (Cambridge, MA: Harvard University Press)
\bigskip
\noindent
Anderson, P. W. (2001)  `Science: A `Dappled World' or a `Seamless Web'?',
{\it Stud. Hist. Phil. Mod. Phys.} {\bf 32}, 487-494; `Reply to Cartwright', 
{\it Stud. Hist. Phil. Mod. Phys.} {\bf 32}, 499-500. 
\bigskip
\noindent
Ax, J. and Kochen, S. (1999) `Extension of Quantum Mechanics to 
Individual Systems', arXiv: quant-ph/9905077. 
\bigskip
\noindent
Bacciagaluppi, G. (2001) `Remarks on Space-time and Locality in Everett's 
Interpretation', talk delivered at the NATO Advanced Research Workshop on 
{\it Modality, Probability, and Bell's Theorems}, Cracow, 19-23 August 2001, 
proceedings to be published by Kluwer Academic Press (NATO Science Series). 
See Pittsburgh PhilSci Archive for the preprint.  
\bigskip
\noindent
Barrett, J. A. (1999)  {The Quantum Mechanics of Minds and Worlds}, 
pp. 227-232.  (Oxford: Oxford University Press)
\bigskip 
\noindent
Bassi, A. and Ghirardi, G. C. (2000) `A general argument against the 
universal validity of the superposition principle', {\it Phys. Lett.} 
{\bf A 275}, 373-381.  
\bigskip
\noindent
Bell, J. S. (1990) `Against ``Measurement''~', in {\it Sixty-Two Years 
of Uncertainty}  (Proceedings of the Erice School, 5-14 August, 1989), 
A. I. Miller, ed. (New York: Plenum Press), pp. 17-31.  This article also  
appears in {\it Physics World}, August 1990, pp. 33-40.  
\bigskip
\noindent
Bohm, D. (1952)  `A suggested Interpretation of Quantum Theory in 
Terms of `Hidden' Variables I, and II', {\it Phys. Rev.} {\bf 85}, 166-193.
\bigskip
\noindent
Brown, H. R. (1986) `The Insolubility Proof of the Quantum Measurement 
Problem', {\it Found. Phys.} {\bf 16}, 857-870. 
\bigskip
\noindent
Bub, J. (1997) {Interpreting the Quantum World}
(Cambridge: Cambridge University Press).  The role of environmental 
decoherence is discussed in Secs. 5.4 and 8.1.   
\bigskip
\noindent
Butterfield, J. (2001)  `Some Worlds of Quantum Theory', to appear 
in the  CTNS/Vatican Observatory volume on Quantum Theory and Divine 
Action, R. Russell et. al., eds.  
\bigskip
\noindent
De Witt, B. S. and Graham, N. (1973) {\it The Many-Worlds Interpretation of 
Quantum Mechanics}  (Princeton: Princeton University Press).
\bigskip
\noindent
Di\'osi, L. (1988) `Continuous Quantum Measurement and It\^o Formalism',
{\it Phys. Lett.} {\bf A 129}, 419-423.  
\bigskip
\noindent
D\"urr D., Goldstein, S. and Zanghi, N. (1992) `Quantum Equilibrium and the 
Origin of Absolute Uncertainty',  {\it Journ. Stat. Phys.} {\bf 67}, 843-907.
\bigskip
\noindent
Fine, A. (1969) `On the general quantum theory of measurement',  
{\it Proc. Cambridge Philos. Soc.} 
{\bf  65}, 111-122.    See also Fine, A. (1970), `Insolubility of the  
Quantum Measurement Problem', {\it Phys. Rev.} {\bf D2}, 2783-2787.
\bigskip 
\noindent 
Ghirardi, G. C., Rimini, A. and Weber, T. (1986) 
`Unified dynamics for microscopic 
and macroscopic systems', {\it Phys. Rev.} {\bf D34}, 470-491.   See also 
Ghirardi, G. C., Pearle, P. and Rimini, A. (1990) `Markov processes 
in Hilbert space and continuous spontaneous 
localization of systems of identical 
particles', {\it Phys. Rev.} {\bf A42}, 78-89.  
\bigskip
\noindent
Gisin, N. (1984) `Quantum Measurements and Stochastic Processes', 
{\it Phys. Rev. Lett.} {\bf 52}, 1657-1660.
\bigskip
\noindent
Gr\"ubl, G. (2002) `The quantum measurement problem enhanced', arXiv: 
quant-ph/0202101; also private email communication to the author.  
\bigskip
\noindent
Harris, R. A. and Stodolsky, L. (1981) `On the time dependence of optical 
activity', {\it J. Chem. Phys.} {\bf 74}, 2145-2155.
\bigskip
\noindent
Joos, E. (1999) `Elements of Environmental Decoherence', in P. Blanchard, 
D. Giulini, E. Joos, C. Kiefer and I.-O. Stamatescu (eds.), 
{\it Decoherence: Theoretical, Experimental, and Conceptual Problems} 
(New York: Springer), pp. 1-17.
\bigskip
\noindent
Joos, E. and Zeh, H. D. (1985)  `The Emergence of Classical Properties 
Through Interaction with the Environment', {\it Z. Phys. B - Condensed 
Matter} {\bf 59}, 223-243.
\bigskip
\noindent
Kent, A. and McElwaine, J. (1997) `Quantum Prediction Algorithms', 
{\it Phys. Rev.} {\bf A55}, 1703-1720.  
\bigskip
\noindent
Pearle, P. (1976) `Reduction of the state vector by a nonlinear Schr\"odinger  
equation', {\it Phys. Rev.}  {\bf D13},  857-868.    
\bigskip
\noindent       
Pearle, P. (1979) `Toward Explaining Why Events Occur', 
{\it Int. Journ. Theor. Phys.} {\bf 18}, 489-518.   
\bigskip
\noindent
Tegmark, M. and Wheeler, J. A. (2001) `100 Years of the Quantum', 
{\it Scientific American} {\bf 284}, 68-75.  
\bigskip
\noindent
Zurek, W. H. (1991) `Decoherence and the Transition from 
Quantum to Classical', {\it Phys. Today} {\bf 44}, No. 10 (October), 36-44.  
\bigskip
\noindent
\bye
\bigskip
\noindent
\bigskip
\noindent
\bigskip
\noindent
\bigskip
\noindent
\bigskip
\noindent
\bigskip
\noindent
\bigskip
\noindent
\bigskip
\noindent
\bigskip
\noindent
\bigskip
\noindent
\bye

\bigskip 
\noindent
\item{[2]} Gisin N 1984 {\it Phys. Rev. Lett.} {\bf 52} 1657
\bigskip
\noindent
\item{[3]}  Ghirardi G C, Rimini A and Weber T 1986 {\it Phys. Rev. D} 
{\bf 34} 470
\bigskip
\noindent
\item{[4]} Ghirardi G C, Pearle P and Rimini A 1990 {\it Phys. Rev. A} 
{\bf 42} 78
\bigskip
\noindent
\item{[5]}  Di\'osi L 1988 {\it J. Phys. A: Math. Gen.} {\bf 21} 2885
\item{~~~}  Di\'osi L 1988 {\it Phys. Lett.} {\bf 129} A 419
\item{~~~}  Di\'osi L 1988 {\it Phys. Lett.} {\bf 132} A 233
\bigskip
\noindent
\item{[6]}  Gisin N 1989 {\it Helv. Phys. Acta.} {\bf 62} 363
\bigskip
\noindent
\item{[7]}  Hughston L P 1996 {\it Proc. Roy. Soc.} A {\bf 452} 953
\bigskip
\noindent
\item{[8]}  Adler S L and Horwitz L P 2000 {\it J. Math. Phys.} {\bf 41} 2485
\bigskip
\noindent
\item{[9]}  Adler S L, Brody D C, Brun T A and Hughston L P 2001 
{J. Phys. A: Math. Gen.} {\bf 34} 8795
\bigskip
\noindent

\bigskip
\noindent
\bigskip
\noindent
\vfill
\eject
\bye